\documentclass[a4paper]{article}
\usepackage{epsfig}
\usepackage{graphicx}  % standard LaTeX graphics tool
\usepackage{color}
\usepackage{lscape}
\usepackage{amsmath}
\date{\today}
 \normalsize

\newcommand{\bmat}{\left(\begin{array}}
\newcommand{\emat}{\end{array}\right)}
\newcommand{\be}{\begin{equation}}
\newcommand{\ee}{\end{equation}}
\newcommand{\bea}{\begin{eqnarray}}
\newcommand{\eea}{\end{eqnarray}}

\def\gtwid{\mathrel{\raise.3ex\hbox{$>$\kern-.75em\lower1ex\hbox{$\sim$}}}}
\def\ltwid{\mathrel{\raise.3ex\hbox{$<$\kern-.75em\lower1ex\hbox{$\sim$}}}}
\def\gev{{\rm \, Ge\kern-0.125em V}}
\def\tev{{\rm \, Te\kern-0.125em V}}

\newcommand{\bd}{\begin{displaymath}}
\newcommand{\ed}{\end{displaymath}}

\def    \be            {\begin{equation}}
\def    \ee            {\end{equation}}
\def    \bea           {\begin{eqnarray}}
\def    \eea           {\end{eqnarray}}
\def\a{\alpha}
\def\b{\beta}
\def\d{\delta}
\def\D{\Delta}
\def\e{\epsilon}
\def\g{\gamma}

\def\l{\lambda}
\def\m{\mu}

\def\t{\tau}
\def\th{\theta}

\def\nn{\nonumber}

\begin{document}
\renewcommand{\thefootnote}{\fnsymbol{footnote}}
%\rightline{IPPP/03/52} \rightline{DCPT/03/104}
\vspace{.3cm}

\title{\Large\bf INTERMEDIATE SCALE DEPENDENCE OF NON-UNIVERSAL GAUGINO MASSES IN SUPERSYMMETRIC SO(10)}

\author
{ \it \bf  Nidal CHAMOUN$^{1,2}$\thanks{nidal.chamoun@hiast.edu.sy}, Chao-Shang HUANG$^{1}$ \thanks{csh@itp.ac.cn}, Chun LIU$^{1}$ \thanks{liuc@mail.itp.ac.cn} and Xiao-Hong WU$^{3}$\thanks{xhwu@ecust.edu.cn} ,
\\ \small$^1$ Key Laboratory of Frontiers in
Theoretical Physics, Institute of Theoretical Physics, \\ \small
Chinese Academy of Sciences, P.O. Box 2735, Beijing 100080, China. \\
\small$^2$ Department of Physics, Higher Institute for Applied
Sciences and Technology,\\\small P.O. Box 31983, Damascus,
Syria.\\ \small$^3$ Institute of Modern Physics,
 East China University of Science and Technology,\\\small
 P.O. Box 532, 130 Meilong Road, Shanghai 200237,  China. }

\maketitle

\begin{center}
\small{\bf Abstract}\\[3mm]
\end{center}
We calculate the dependence on intermediate scale of the gaugino
mass ratios upon breaking of $SO(10)$ into the SM via an
intermediate group $H$. We see that the ratios change
significantly when the intermediate scale is low (say, $10^8$ GeV
or $1$ Tev) compared to the case when the two breakings occur at
the same scale.

\vspace{0.5cm}

\begin{minipage}[h]{14.0cm}
\end{minipage}

\hrule \vskip 0.3cm
\section{Introduction}
With the Large Hadron Collider (LHC) having started to operate, the high
energy community is expanding focus in the study and search of
new physics beyond the standard model (SM).

Grand unification theories (GUTs) are models of the most promising
ones for this new physics \cite{langacker}. However, supersymmetry
(SUSY) is necessary here to make the huge hierarchy between the GUT
scale and the electroweak scale stable under radiative
corrections \cite{martin-primer}. In this regard, SUSY SO(10) is an appealing candidate for
realistic GUTs \cite{mohapatra-book}. Universal boundary conditions for gaugino masses,
as well as other soft terms, at the high scale (the unification
scale or Plank scale) are adopted in the the setting of the minimal supergravity (mSUGRA) or the constrained minimal
supersymmetric standard model CMSSM \cite{cmssm}.
If the discrepancy between the SM theoretical predictions and the experimental
determinations of $(g-2)$ is confirmed at the $3$-sigma level, this could
be interpreted as strong evidence against the CMSSM \cite{g-2}. Non-universal
gaugino masses may arise in supergravity models in which a
non-minimal gauge field kinetic term is induced by the
SUSY-breaking vacuum expectation value (vev) of a chiral
superfield that is charged under the GUT group G \cite{anderson}. The
non-universal gaugino masses resulting from SUSY-breaking vevs of non-singlet chiral superfields,
for $G=SU(5)$, $SO(10)$ and $E_6$, and
their phenomenological implications have been
investigated in~\cite{chamoun,chakra1,martin,chakra2}.

If the grand unification group $G$ is large enough, like
$SO(10)$ or $E_6$, then there are more than one breaking chain from $G$ down to the SM.
It is natural here to assume that there exist multi
intermediate mass scales in the breaking chain. It has been found
that when extrapolating the coupling strengths to very high
energies, they tend to converge in the non-SUSY $SO(10)$ provided one introduces two
new intermediate energy scales, whereas they do not meet at one point in the
absence of intermediate energy scale~\cite{alp}. A systematic study of the constraints of
gauge unification on intermediate mass scales in non-SUSY $SO(10)$ scenarios was recently
discussed in~\cite{bertolini}.

The possibility of the existence of intermediate scales is an
important issue for supersymmetric unification. The success of the minimal supersymmetric standard model
MSSM couplings unification~\cite{giunti} favors a single GUT
scale, and the intermediate scales cannot be too far from the GUT
scale. However, recent studies show that in GUTs with large number
of fields renormalization effects significantly modify the scale
at which quantum gravity becomes strong and this in turn can
modify the boundary conditions for coupling unification if higher
dimensional operators induced by gravity are taken into
consideration~\cite{calmet}. In GUT model building, the so called magic
fields can be used to fix the gauge coupling unification in certain
two-step breakings of the unified group~\cite{cfrz}. It has been
pointed out that any choice of three options - threshold
corrections due to the mass spectrum near the unification scale,
gravity induced non-renormalizable operators near the Plank scale,
or presence of additional light Higgs multiplets - can permit
unification with a lower intermediate scale~\cite{majee}. This
unification with distinct energy scales yields right handed
neutrino masses in the range ($10^8-10^{13}$ GeV) relevant for
leptogenesis~\cite{luty}, perhaps even reaching the TeV
region~\cite{majee}.

In the previous studies~\cite{chamoun,chakra1,martin,chakra2} on
non-universal gaugino masses in SUSY-$SO(10)$ one assumed for
simplicity that there was no intermediate scales between $M_{GUT}$
and $M_S$ (the SUSY scale$\sim 1TeV$) or the electro-weak scale
$M_{EW}$.
In this paper, we study in detail the intermediate scale dependence
of the non-universal gaugino masses.

The starting point is to consider a chiral superfield (`Higgs'
field) $\Phi$ transforming under the gauge group $G=SO(10)$ in an
irrep $R$ lying in the symmetric product of two adjoints \footnote{All group theory considerations can be found in the reviw article \cite{slansky81}.}: \bea
\label{45-sym-prod} ({\bf 45}\times{\bf 45})_{symmetric}&=&{\bf
1}+{\bf 54}+{\bf 210}+{\bf 770} \eea If R is G non-singlet and
$\Phi$ takes a vev (vacuum expectation value) spontaneously
breaking $G$ into a subgroup $H$ containing the SM, then it can
produce a gauge non-singlet contribution to the $H$-gaugino mass
matrix \cite{ellis85} \bea M_{\a,\b}&=& \eta_\a \d_{\a\b} \langle
\Phi \rangle \eea where the discrete $\eta_\a$'s are determined by
$R$ and $H$.

Here, we make two basic assumptions. The first one is to omit the
`possible' situation of a linear combination of the above irreps
and to consider the dominant contribution to the gaugino masses
coming from one of the non-singlet $F$-components. The second
assumption is that $SO(10)$ gauge symmetry group is broken down at
GUT scale $M_{GUT}$ into an intermediate group $H$ which, in turn,
breaks down to the SM at some intermediate scale $M_{HB}$. In the
case of several intermediate symmetry breakings one can assume various intermediate
scales, for which case it is straightforward to generalize our method.

We insist on $H$ being the gauge symmetry group in the range from
$M_{HB}$ to $M_{GUT}$. Thus, only the $F$-component of the field
$\Phi$ which is neutral with respect to $H$ can acquire a vev
yielding gaugino masses. Depending on the breaking chain one
follows down to the SM, ratios of gaugino masses $M_a$'s are
dependent of $M_{HB}$ and are determined purely by group theoretical factors
only if $M_{HB}=M_{GUT}$. In fact, the functional dependence on
$M_{HB}$ of the gaugino mass ratios can not be deduced from their values obtained in the
case of $M_{HB}=M_{GUT}$ by mere renormalization group (RG)
running, and one has to consider carefully the normalization of the group
generators and the mixing of the abelian $U(1)$'s necessary to get
the dependence of the $U(1)_Y$'s gaugino mass on the intermediate
scale.

Whereas in ref.\cite{chamoun} we considered only low dimensional irreps $\bf {54},~{\bf
210}$,  we extend here our analysis to include all three
non-singlet irreps. Moreover, there were some errors in the results of ref.\cite{chamoun},
which upon being corrected agree now with the conclusions of
\cite{chakra1,martin,chakra2} when  $M_{HB}=M_{GUT}$.

The plan of this paper is as follows. In section 2, we consider
the first step of breaking, from $GUT=SO(10)$ to the intermediate
group $H$, and calculate the $H$-gaugino mass ratios at the GUT scale $M_{GUT}$, for the three
cases $H=G_{422}\equiv SU(4)_C \times SU(2)_L \times
SU(2)_R,H=SO(3)\times SO(7)$ and $H=H_{51}\equiv SU(5) \times
U(1)_X$, depending on the specific irreps in
Eq.(\ref{45-sym-prod}). We investigate the second step of the
breaking, from the intermediate group $H$ to the SM group in section 3, and compute
the MSSM gaugino masses in terms of the $H$-gaugino masses at the intermediate breaking scale $M_{HB}$.
 Taking the RG running from $M_{GUT}$ to $M_{HB}$ into consideration, we compute in section 4 the
 MSSM gaugino mass ratios at $M_{HB}$. We also state in this section the particle content of the model in each case,
and calculate the beta function coefficients necessary
for the RG running. In section 5, we summarize the results in form
of a table, where we compare numerically the case of two breaking
scales with the case of one breaking scale, and present our
conclusions.

\section{From $GUT=SO(10)$ to the intermediate group $H$}
Here we discuss the different ways in which one can break the
$GUT$-group $SO(10)$ depending on the Higgs irrep one uses. As
noted earlier, three irreps can be used (see
Eq.\ref{45-sym-prod}): $\bf {54,210}$ and $\bf 770$.

\subsection{The irrep $\bf 54$}
If an irrep ${\bf 54}$ is used then the branching rules for
$SO(10)$ tell us it can be broken into several subgroups (e.g.
$H=G_{422}, H=SU(2)\times SO(7), H=SO(9)$). The choice $H=SO(9)$
leads to universal gaugino masses whereas the other two possible
chains are more interesting.

\subsubsection{$H=G_{422}$}
The ${\bf 54}$ irrep can be represented as a traceless and
symmetric $10 \times 10$ matrix which takes the vev: \bea <{\bf
54}>&=& v~
Diag(\underbrace{2,\dots,2}_{6},\underbrace{-3,\ldots,-3}_{4})
\eea with the indices $1,\ldots 6$ corresponding to $SO(6)\simeq
SU(4)_C$ while those of $7,\ldots 0$ (henceforth $0$ means $10$) correspond to $SO(4)\simeq
SU(2)_L\times SU(2)_R$.

This implies that at $M_{GUT}$-scale we have: \bea \label{54G422}
\left.\frac{M_L}{M_4}\right|_{M_{GUT}}=\left.\frac{M_R}{M_4}\right|_{M_{GUT}}&=&
-\frac{3}{2}\eea

\subsubsection{$H=SU(2)_L\times SO(7)$}
The first breaking is achieved by giving a vev to the irrep ${{\bf
54}}$ \bea <{\bf 54}>&=& v
Diag(\underbrace{7/3,\dots,7/3}_{3},\underbrace{-1,\ldots,-1}_{7})
\eea where the indices $1$,$2$,$3$ correspond to $SO(3)\simeq
SU(2)_L$ and $4,\ldots 0$ correspond to SO(7). This gives at
$M_{GUT}$-scale \bea \label{54s03so7} \left.\frac{M_L}{M_7}\right|_{M_{GUT}}&=&
-\frac{7}{3}\eea

\subsection{The irrep $\bf 210$}
This irrep can be represented by a $4^{th}$-rank totally
antisymmetric tensor $\Delta_{abcd}$. It can break SO(10) in
different ways, of which we consider two.
\subsubsection{$H=G_{422}$}
The first breaking from $SO(10)$ to $H$ is achieved when the only
non-zero vev is $<\Delta_{abcd}>=v\epsilon_{7890}=v$ \cite{aulakh}
where ($a,b,c,d\in\{1,\ldots,0\}$). This leads to the mass term:
\bea {\cal L}_{\text{mass}} \propto <\Delta_{abcd}> \l^a_b \l^c_d %&=& v \l^7_8 \l^9_0 \\
&=& \frac{v}{4}[(\l^7_8+\l^9_0)^2-(\l^7_8-\l^9_0)^2]
\label{210mass422}\eea  As the indices ($1,\ldots,6$) which
correspond to $SO(6)$ do not appear in the mass term then we have
\bea \label{0} \left.M_4\right|_{M_{GUT}} &=& 0\eea We can take the gauginos
$\l_{2L},\l_{2R}$ corresponding to $SU(2)_L,SU(2)_R$ as being
proportioanl to the `bracketed' combinations of $\l^7_8$ and
$\l^9_0$ in Eq.(\ref{210mass422}), and thus we get:
 \bea \label{-1} \left.\frac{M_{R}}{M_{L}}\right|_{M_{GUT}} &=& -1.\eea

\subsubsection{$H=H_{51}$}
This breaking from $SO(10)$ occurs when \cite{he90}: \bea
 \Delta_{1234}&=&\Delta_{1256}=\Delta_{1278}=\Delta_{1290}=
\Delta_{3456} =\Delta_{3478}\nonumber\\
&=&\Delta_{3490}=\Delta_{5678}=\Delta_{5690}= \Delta_{7890}=v.
\eea For the $H=H_{51}$-case, we adopt the convention of restricting the use of indices to the
$SU(5)$-indices in order to express only the $SU(5)\times U(1)_X$ gauginos
amongst the $SO(10)$-ones. In fact, the branching rule \bea
\label{10branching rule} {\bf 10} &\overset{SO(10)\supset
SU(5)\times U(1)_X}{=}& ({\bf 5})_{2} +
({\bf \overline{5}})_{-2} \eea %in Eq.(\ref{10branching rule})
allows us to use the indices: \bea \label{indices}
i=\tilde{a}+\tilde{b}\equiv a+\bar{b} &\text{with}&
i=\in\{1,\ldots,0\};\tilde{a}\equiv
2a-1\in\{1,3,5,7,9\};\tilde{b}\equiv 2\bar{b}\in\{2,4,6,8,0\}\eea
and so we have the $SU(5)$-indices
($a=1,\ldots,5;\bar{b}=\bar{1},\ldots,\bar{5}$) written usually as
an upper index for `$a$' and a lower index for `${b}$' (omitting
the `bar' of $\bar{b}$). With this, we write the $SO(10)$ adjoint
irrep $\l^{(2a-1)(2b)}$ as $\l^a_b$ using the $H_{51}$ indices
($a,b=1,\ldots,5$).

We know that the only way to get a $4^{\text{th}}$-rank totally
antisymmetric tensor invariant under $SU(5)$ is by considering:
\bea \e^{abefg}\e_{cdefg} &=& \d^a_c\d^b_d - \d^a_d \d^b_c\eea
($a,b,c,d,e,f,g=1,\ldots,5$) and thus the $H_{51}$-singlet takes
on the invariant form \bea <\D^{ab}_{cd}> &=& v
\e^{abefg}\e_{cdefg} \eea The gaugino mass term becomes \bea
<\D^{ab}_{cd}> \l^c_a \l^d_b &\propto& - \widehat{\l^c_a}
\widehat{\l^a_c} + \frac{4}{5} (\l^b_b)^2 = - (\widehat{\l^c_a})^2
+ 4 (\l)^2\eea where the `traceless' $SU(5)$-gaugino
$\widehat{\l^c_a}$ and the $U(1)_X$-gaugino $\l$ are defined as
usual by: \bea
\label{labH51}\widehat{\l^a_b} &=& \l^a_b - \frac{1}{5} \d^a_b
\l^c_c \\ \label{lH51} \l &=& \frac{1}{\sqrt{5}}
 \l^c_c\eea We get at $M_{GUT}$ the ratio: \bea \label{-4}
\left.\frac{M_X}{M_5}\right|_{M_{GUT}}&=& -4\eea

\subsection{The irrep $\bf 770$}
This irrep can be represented by a traceless $4^{\text{th}}$-rank
tensor $\phi^{ij,kl}$ with symmetrized and anti-symmetrized
indices in the combinations corresponding to the Young diagram
with two rows and columns. It can break $SO(10)$ in three ways.

\subsubsection{$H=G_{422}$}
Here, since we have the branching rule: \bea  {\bf 10}
&\overset{SO(10)\supset SO(6)\times SO(4)}{=}& ({\bf 1, 4})+({\bf
6, 1})\overset{SO(10)\supset SU(4)\times SU(2) \times
SU(2)}{\equiv}({\bf 1, 2, 2})+({\bf 6},{\bf 1},{\bf 1}), \eea we
can set $\phi^a=\phi^\a +\phi^i$ with
$a=1,2,...,0;\;\a=1,...,6;\;i=7,...,0$. When the scalar components
of $\phi^{ab,cd}$,
  corresponding to the singlet ($\bf1,1$) of ${\bf 770}$ under
$SO(10) \supset SO(6)\times SO(4)$, acquire a non-zero vev, then
the tensor structure impose the form: \bea
<\phi^{\a\b,\g\d}> &=& v(\d^{\a\b}\d^{\g\d}-\d^{\a\d}\d^{\b\g}) \nn\\
<\phi^{ij,kl}> &=& sv(\d^{ij}\d^{kl}-\d^{il}\d^{jk}) \nn\\
<\phi^{\a\b,ij}> &=& s^\prime v\d^{\a\b}\d^{ij} \label{tensor
structure} \eea ($\a,\b,\g,\d=1,\ldots,6;i,j,k,l=7,\ldots,0$).
Forcing the tensors $\phi^{aa\g\d}$ and $\phi^{aaij}$ to be traceless would imply  $s^\prime = -\frac{5}{4}$
and $s= \frac{5}{2}$, and so one gets a mass term:
\bea {\cal L}_{\text{mass}} &=& \phi^{\a\b\g\d} \l_{\a\b}\l_{\g\d}+ \phi^{ijkl} \l_{ij}\l_{kl}\\
&=& -v (\l_{\a\b})^2-sv(\l_{ij})^2\eea The $\l_{\a\b}$'s
correspond to $SO(6)$-gauginos whereas $\l_{ij}$'s correspond to
$SO(4)$-gauginos, whence we get at $M_{GUT}$-scale the ratios:
\bea \label{5/2} \left. \frac{M_{L}}{M_{R}} \right |_{M_{GUT}} = 1 &,&
\left.\frac{M_{R}}{M_4}\right |_{M_{GUT}}= \frac{5}{2}\eea
\subsubsection{$H=SO(3)
\times SO(7) \simeq SU(2)_L\times SO(7)$}
 Again, the branching rule:
\bea {\bf 10} &\overset{SO(10) \supset H_{51}}{\Large {=}}& ({\bf
3, 1})+({\bf 1, 7}) \eea enables us to set $\phi^a=\phi^\a
+\phi^i$ with $a=1,\ldots,0;\;\a=1,\ldots,7;\;i=8,9,0$. In the
same way as in the case of $H=G_{422}$, when the scalar components
of $\phi^{ab,cd}$,
  corresponding to the singlet ($\bf1,1$) of ${\bf 770}$ under
$SO(10) \supset SO(3)\times SO(7)$, acquire a non-zero vev then we
have the same tensor structures as in Eqs.(\ref{tensor
structure}). Forcing the traces $\phi^{aa\g\d}$ and $\phi^{aaij}$
to vanish would imply $s^\prime = -2$ and $s=7$. Substituting in
the Lagrangian gaugino mass term gives now at $M_{GUT}$ the
ratios: \bea \label{7} \left. \frac{M_{L}}{M_{R}} \right |_{M_{GUT}} = 1 &,&
\left.\frac{M_{R}}{M_4}\right |_{M_{GUT}}= 7 \eea
 \subsubsection{$H=H_{51}$}
Again, using the branching rule in Eq.(\ref{10branching rule}), we
can take $\phi^a=\phi^i +\phi^{\bar k}\equiv \phi^j+\phi_l$ with
$a=1,\ldots,0; i=1,3,5,7,9\equiv 2j-1;\bar k=2,4,6,8,0\equiv 2l$
($j,l=1,\ldots,5$ are the $\bf 5$ and $\bf \bar 5$ indices
respectively). When the traceless $4^{\text{th}}$-rank tensor
$\phi^{ab,cd}$ scalar fields, corresponding to the singlet
($\bf1,1$) of ${\bf 770}$ under $SO(10) \supset H_{51}$, have a
non-zero vev, then we have the following tensor structures: \bea
\phi^{ab,cd}&=&\phi^{ij,kl}+\phi^{ij,k}_{l}+\phi^{ij}_{kl}+
\phi^{kl}_{ij} + \phi^{i}_{j,kl} +
\phi_{ij,kl} \label{tensor structure first equation}\\
<\phi^{ij,kl}>&=&v_1(\delta^{ij}\delta^{kl}-\delta^{kj}\delta^{il}) \nn\\
<\phi^{ij,k}_{l}>&=&v_2\delta^{ij}\delta^k_l \nn \\
<\phi^{ij}_{kl}>&=&v_3(\delta^{ij}\delta_{kl}
+\delta^i_k\delta^j_l+\delta^i_l\delta^j_k) \label{tensor
structure H51-770 } \eea
($a,b,c,d=1,\ldots,0;i,j,k,l=1,\ldots,5$). Note that since $SU(5)$
is the only maximal non-abelian subgroup in $H_{51}$ then all the
vevs above are equal $v_1=v_2=v_3=v$. We note also that the
contribution to the gaugino mass from the last three terms in
Eq.(\ref{tensor structure first equation}) is equal to that coming
from the first three terms, and thus we can limit the computation
to these latter terms to get the mass term: \bea \langle
\phi^{ab,cd}\rangle \l_{ab}\l_{cd} &=& v [\widehat{\l^j_l}
\widehat{\l^l_j} + 16 \l^2]\eea where the expressions of the
`traceless' $SU(5)$-gaugino $\widehat{\l^j_l}$ and the
$U(1)_X$-gaugino $\l$ are taken from Eqs.(\ref{labH51}) and
(\ref{lH51}). We get at $M_{GUT}$ the ratio: \bea \label{ratio16}
\left. \frac{M_X}{M_5} \right|_{M_{GUT}}&=& 16\eea

\section{From the intermediate group to the SM}

We discuss here the second stage of the breaking from $H$ into the $SM$. We note that
in some cases there are more than one $U(1)$-group, and we need to consider the mixing of these
 $U(1)$'s in order to get the $U(1)_Y$ of the SM. The method is
standard and we work it out case by case.

\subsection{$H=G_{422}\equiv SU(4)_C
\times SU(2)_L \times SU(2)_R \rightarrow SM\equiv SU(3)_C \times
SU(2)_L \times U(1)_Y $} %One can fulfill the breaking $SO(10)
%\rightarrow H=G_{422}$ by using anyone of the $SO(10)$-irreps:
%${\bf 54},{\bf 210}$ and ${\bf 770}$.
The Higgs field responsible
for the breaking $SU(4)_C \times SU(2)_R \rightarrow SU(3)_C
\times U(1)_Y$ can be taken to include the irrep $({\bf 4},{\bf
2})$ of the group $SU(4)_C \times SU(2)_R$: \bea \Phi &=&
\varphi^a \bigotimes \varphi^r : a\in\{1,2,3,4\},r\in\{1,2\}\eea
We can choose $\Phi$ to be in the spinor irrep of $SO(10)$ since
we have the branching rule: \bea {\bf 16} &\overset{SO(10)\supset
G_{422}}{=}& (\bf{4,2,1})+(\bf{\overline{4},1,2})\eea and we can
write the covariant derivative terms related to the $SU(4)_C
\times SU(2)_R$ group in the form: \bea {\cal{D}}_\m \Phi &=&
\partial_\m \Phi - ig_4 \frac{T^b}{2}A^b \varphi^a - ig_R
\frac{\t^s}{2}B^s \varphi^r \eea where
$T^b$($b\in\{1,\ldots,15\}$) are the $4\times4$ generalized
Gellman matrices for $SU(4)$ with the standard normalization
$Tr(\frac{T^a}{2}\frac{T^b}{2})=\frac{1}{2}\d^{ab}$,
$\t^r$($r\in\{1,2,3\}$) are the $2\times2$ Pauli matrices
satisfying $Tr(\frac{\t^r}{2}\frac{\t^s}{2})=\frac{1}{2}\d^{rs}$.

In  order to break $SU(4)_C$ to $SU(3)_C \times U(1)$, and
$SU(2)_R$ to $U(1)^\prime$, the Higgs fields take the vevs: \bea
\langle \varphi^a \rangle = v_1 \d^{a4} &,& \langle \varphi^r
\rangle = v_2 \d^{r1} \eea Since both $\varphi^a$ and $\varphi^r$
originate from the same $\Phi$, the spinor irrep in $SO(10)$ which
under $SO(10)\supset SM$ has the component $({\bf{1,1}})_0$, then
the two vevs are equal: $v_1=v_2=v$. Concentrating on the mixing
of the $U(1)$ from $SU(4)_C$ and the other $U(1)^\prime$ from
$SU(2)_R$, we note that the corresponding $A^{15}$ and $B^3$
components will mix together, and thus we obtain the neutral gauge
boson mass terms in the form : \bea \langle D_\m\Phi \rangle
\langle D_\m\Phi \rangle ^+ &=& \frac{v^2}{4} \left(  \sqrt{\frac
{3}{2}} g_4 A^{15}-g_R B^3\right)^2 \eea  This quadratic form in
the fields $B^3$ and $A^{15}$ has a zero eigenvalue whose
corresponding eigenstate can be identified as the massless
$U(1)_Y$ gauge boson $E$. By diagonalizing the corresponding mass
matrix we obtain the two physical vector bosons: the massless
gauge boson $E$ , and the orthogonal combination $F$ corresponding
to a massive vector boson: \bea F&=& \cos\theta A^{15}-\sin\theta
B^3 \nn \\E&=&\sin\theta A^{15}+\cos\theta B^3\label{mix1} \eea
where \bea \label{theta1}
\cos\theta=\frac{\sqrt{\frac{3}{2}}g_4}{c},~~~~~~~\sin\theta=\frac{g_R}{c}
&:& c^2=g_R^2+\frac{3}{2}g_4^2 \eea

It is convenient \cite{langacker} to define the
$4\times4$ ($2\times2$) matrix $\bf{A}$ ($\bf{B}$) as follows \bea
{\bf{A}}=\frac{T^b A^b}{\sqrt{2}}\text{ with
}A^a_b\equiv(\bf{A})_{ab}&,& {\bf{B}}=\frac{\t^r
B^r}{\sqrt{2}}\text{ with
}B^r_s\equiv(\bf{B})_{rs}\label{nota}\eea which leads to \bea
\label{a44anda15} A^4_4=-\frac{\sqrt{3}}{2}A^{15} &,&
B^1_1=\frac{B^3}{\sqrt{2}}\label{rel}\eea

In the notation of Eq. (\ref{nota}), the gaugino fields which lie in the same
supermultiplet as the gauge fields $A^a_b$ of the $SU(4)_C$ group are
denoted by $\l^a_b$ ($a,b=1,\ldots,4$ with $\l^a_a=0$), whereas we denote the gaugino fields of
the $SU(2)_{L,R}$ group by ${\l^r_{s}}_{L,R}$
($r,s=1,2$ with $\l^r_r=0$). Then the gaugino
mass term in the $G_{422}$ group is: \bea
\label{g422massterm} \text{mass term} &=& M_4 \l^a_b \l^b_a + M_L
{\l^r_s}_L {\l^s_r}_L + M_R {\l^r_s}_R {\l^s_r}_R \nn\\ &=& M_4
\widehat{\l^\a_\b} \widehat{\l^\b_\a} + \frac{4}{3}M_4 (\l^4_4)^2
+ M_L {\l^r_s}_L {\l^s_r}_L + 2 M_R ({\l^1_1}_R)^2 + \ldots\eea
where $\widehat{\l^\a_\b} = \l^\a_\b - \frac{1}{3}
\d^\a_\b \l^\g_\g$ ($\a,\b=1,2,3$) are the $SU(3)_C$ gaugino fields
and `$\ldots$' denote the terms which do not contribute to the MSSM
gaugino masses.

Since the gaugino mixing should proceed in the same way as that for the gauge
fields lying in the same supermultiplet, then Eqs. (\ref{mix1} and
\ref{rel}) lead `by supersymmetry' to: \bea \label{l44}\l^4_4 &=&
-\frac{\sqrt{3}}{2}(\sin\theta \l+\cos\theta \widetilde{\l})
\\ \label{l11} {\l^1_1}_R &=&\frac{1}{\sqrt{2}}(\cos\theta \l-\sin\theta \widetilde{\l})
\eea where $\l$ is the gaugino field lying in the same
supermultiplet as the $U(1)_Y$ gauge field $E$, whereas
$\widetilde{\l}$ is the superpartner of the massive vector boson
$F$.

It follows from Eq.(\ref{g422massterm}) that at the intermediate
scale $M_{HB}$ we have: \bea \label{mass_rel_g422_first} \left.
M_3\right|_{M_{HB}}=\left. M_4\right|_{M_{HB}} &,&
\left.M_2\right|_{M_{HB}}=\left. M_L\right|_{M_{HB}}\eea As to the
mass term corresponding to $U(1)_Y$, then substituting
Eqs.(\ref{l44} and \ref{l11}) into Eq.(\ref{g422massterm}) leads
to: \bea \label{mass_rel_g422_second} \left. M_1\right|_{M_{HB}} =
\sin^2\theta M_4+\cos^2\theta M_R  &=& \left.
\frac{2g_R^2M_4+3g_4^2M_R}{3g_4^2+2g_R^2}\right|_{M_{HB}} \eea

To summarize, we have used an $SO(10)$--${\bf 16}$
irrep Higgs field to break $G_{422}$ into the SM when its neutral
component $({\bf 1, 1})_0$ under SM develops a vev. The gauge
supermultiplets ${\bf 45}$ of $SO(10)$ would also be decomposed
having under $G_{422}$ the components $({\bf 15,1,1})$ and $({\bf
1,1, 3})$representing respectively the generators of $SU(4)$ and
$SU(2)_R$. In the breaking from $G_{422}$ to SM, each of the
latter generators would have a singlet $({\bf 1, 1})_0$ part and
one needs to identify the weak hypercharge $Y$ generator as a
linear combination of these $({\bf 1, 1})_0$ parts. With this, we
could determine the $U(1)_Y$ gaugino in terms of the gauginos and
coupling constants $g_4$,
$g_R$ corresponding to $SU(4)_C$ and $SU(2)_R$.

\subsection{$H\equiv SO(3)\times SO(7) \rightarrow H^\prime \equiv SU(2)_L \times SO(6) =
SU(2)_L \times SU(4) \rightarrow SM\equiv SU(3)_C \times SU(2)_L
\times U(1)_Y $}
 As we have discussed, one can use the irreps $\bf 54$ or
$\bf 770$ to carry out the breaking $SO(10)\rightarrow SO(3)\times
SO(7) \equiv SU(2)_L \times SO(7)$. As pointed out in
\cite{martin}, the $SU(2)\times SO(7)$ can not be reconciled with
the chiral fermion content of the SM. However, as was noticed in
\cite{ross}, this case produces non-trivial mass ratios with
interesting phenomenology, and we may still consider it since we
are not involved in the model building. Thus, until the
identification of a feasible model with masses in this region, we
include the examination of this case in our study.

Now, the $SO(7)$ is
broken at $M_{GUT}$ to $SO(6)\simeq SU(4)$ which in turn is broken
to $SU(3)_C\times U(1)_Y$ at $M_{HB}$. One can not use the $SU(4)$-$\bf 4$ irrep to achieve this breaking since its branching rule is: \bea {\bf 4} &\overset{SU(4)\supset SU(3)\times U(1)}{=}& {\bf 1}_3+{\bf 3}_{-1}\eea whereas the `next simple'
$SU(4)$-$\bf 15$ irrep can carry out this breaking having the branching rule:
\bea{\bf 15} &\overset{SU(4)\supset SU(3)\times U(1)}{=}& {\bf 1}_0+{\bf 3}_{-4} + {\bf \overline{3}}_{4}+ {\bf 8}_{0}\eea
Thus, the Higgs field $\Phi$ responsible for the breaking $SU(4)\rightarrow SU(3)_C\times U(1)_Y$ should include the
 $SU(4)$-$\bf 15$ irrep, and the simplest choice is the $\bf 45$ irrep of $SO(10)$ having the branching rules:
 \bea {\bf 45}&\overset{SO(10)\supset SO(3)\times S(7)}{=}& ({\bf 3,1}) + ({\bf 1,21}) + ({\bf 3,7}) \\
  {\bf 21}&\overset{SO(7)\supset SO(6)}{=}& {\bf 15} + {\bf 6} \eea

  The ($SO(7)$) gaugino mass term in the Lagrangian is \bea {\cal L}^{SO(7)}_{\text{mass}} &=&
   M_7 \l^{[a,b]}\l_{[a,b]}= M_7 \l^{[\a,\b]}\l_{[\a,\b]}+M_7 \l^{[7,\a]}\l_{[7,\a]}\eea
  where $a,b=1,\ldots,7;\a,\b=1,\ldots,6$. Note that the $\l^{[7,\a]}$ does not represent
 the superpartner of a gauge field in $SO(6) = SU(4)$, and thus, using the $SU(4)$ indices, the mass term of the $SU(4) \times SU(2)_L$ is
 \bea {\cal L}_{\text{mass}} &=&
  M^{\prime}_4  \l^i_j \l^j_i + M_L \l^r_s \l^s_r \nn \\ &=& M^{\prime}_4  \l^\a_\b \l^\b_\a + M^{\prime}_4 \l^4_4 \l^4_4 + M_L \l^r_s \l^s_r
  + \ldots  \label{7mass}
  \eea where $i,j = 1,\ldots,4$ (with $\l^i_i = 0$); $r,s=1,2$ (with $\l^r_r=0$);
  $\a,\b = 1,2,3$ and the `$\ldots$' represent the terms which do not contribute to the gaugino masses:
  $M_L$ for $SU(2)$ and $M^{\prime}_4$ for $SU(4)$ satisfying \bea \label{m4m7} \left. \frac{M^{\prime}_4}{M_7}\right|_{M_{GUT}} = 1\eea

We introduce in the same way as we did before,
the `traceless' $SU(3)$-gauginos: $\widehat{\l^\a_\b}
= \l^\a_\b -
  \frac{1}{3} \d^\a_\b \l^\g_\g$, and the `squared' $U(1)_Y$ gaugino field $\l^2=\frac{1}{3} (\l^\g_\g)^2 + (\l^4_4)^2$.
Eq.(\ref{7mass}) reduces then to
  \bea \label{massso7}{\cal L}_{\text{mass}} &=&
  M^{\prime}_4  \widehat{\l^\a_\b} \widehat{\l^\b_\a} + M^{\prime}_4 \l^2 +  M_L \l^r_s \l^s_r \eea
 Therefore, we have at $M_{HB}$, the scale where the breaking of the intermediate group $H^\prime$ takes
  place, the relations:

  \bea  \label{mass_rel_so7} \left.M_1\right|_{M_{HB}}=\left.M_3\right|_{M_{HB}}=\left.M^{\prime}_4\right|_{M_{HB}} &,&
  \left.M_2\right|_{M_{HB}}=\left.M_L\right|_{M_{HB}} \eea

\subsection{$H=H_{51}\equiv SU(5)\times U(1)_X\rightarrow SM\equiv SU(3)_C
\times SU(2)_L \times U(1)_Y $} In order to break $SU(5)$ to
$SU(3)_C \times SU(2)_L \times U(1)_Z$, one can use the
$(SU_5)$-$10$-irrep with
 the branching rule:
 \bea \label{10}{\bf 10} &\overset{SU(5)\supset SU(3)_C\times SU(2)_L\times U(1)_Z}{=}& ({\bf 3^*,1})_{-\frac{2}{3}} + ({\bf 3,2})_{\frac{1}{6}} + ({\bf 1,1})_1\eea
 Thus, the Higgs field $\Phi$ responsible for the breaking $H_{51}\rightarrow SM$ can be taken in the $(SO10)$-$16$-irrep
 having the branching rule: \bea \label{16} {\bf 16} &\overset{SO(10)\supset SU(5)\times U(1)_X}{=}& {\bf 10}_1 + {\bf \bar{5}}_{-3}+{\bf 1}_5 \eea

 The conventions in the above two branching rules are consistent with the $U(1)_Z$-generator in $SU(5)$ given by:
 \bea \label{Z} Z &=& diag(-1/3,-1/3,-1/3,1/2,1/2)\eea
 and we have
 an unbroken hypercharge~\cite{barr}: \bea \frac{Y}{2} = \frac{1}{5} (X-Z)\eea
As it is well known, one needs to define the `properly normalized'
$U(1)_Z$-generator to be:
  \bea \label{L_Z} L_Z &=& \sqrt{\frac{3}{5}} Z\eea so that $Tr(L_Z)^2 = \frac{1}{2}$.
Similarly, we define the `properly normalized' $U(1)_X$-generator
to be: \bea \label{L_X} L_X &=& \sqrt{\frac{1}{40}} X \eea such
that $Tr_{{\bf 10}}(L_X)^2 = 1$, since we should have $Tr_{{\bf
10}}(M_{ij}M_{i^\prime
j^\prime})=1\delta_{ii^\prime}\delta_{jj^\prime}$ where $M_{ij}$
is the $SO(10)$ generator and ${\bf 10}$ is the defining
(vector) irrep of $SO(10)$, and that the branching rule  \bea \label{10branching rule1} {\bf 10}
&\overset{SO(10)\supset SU(5)\times U(1)_X}{=}& ({\bf 5})_{2} +
({\bf \overline{5}})_{-2} \eea implies $ Tr_{{\bf 10}}(X^2)=40$.

We now come to the mixing of the two $U(1)$'s, which means we study how $U(1)_Z\times U(1)_X$
breaks into $U(1)_Y$. When the Higgs field corresponding to
the $(1,1)$ component of Eq. (\ref{10}), with $Z$- and $X$-charges equal to one
and represented by a $5\times 5$ antisymmetric tensor $\phi^{ab}$, takes a vev such that the only non-zero elements are:
 \bea <\phi^{45}> = - <\phi^{54}>&=& v \eea
 we get a mass term
\bea \label{mass H51} {\cal L}_{\text{mass}}
&=&
v^2 \left( g_5 \sqrt{\frac{3}{5}}A^Z_\m + \frac{g_X}{\sqrt{40}}
B^X_\m\right)^2\eea where $A^Z$ and $B^X$ are the $U(1)_Z$ and
$U(1)_X$ gauge fields, respectively.

By diaganolizing the mass matrix corresponding to the above quadratic form, we get a massive
$U(1)_Y$-neutral vector boson field $B_\m$ and a massless
$U(1)_Y$-gauge field $A_\m$ given by: \bea B&=& \cos\psi
A^Z-\sin\psi B^X \nn \\ A&=&\sin\psi A^Z+\cos\psi B^X
\label{mix2}\eea where \bea
\cos\psi=\frac{\sqrt{3}g_5}{c},~~~~~~~\sin\psi=-\frac{g_X}{\sqrt{8}c}
&:& c^2=3 g_5^2+\frac{g_X^2}{8}\eea

Let $\l,\widetilde{Z}$ be the superpartners of $B^X$, $A^Z$
respectively, and call $\widetilde{X}$ the superpartner of the massive $B$,
whereas we denote the superpartner of the massless $A$, that is the
$U(1)_Y$ gaugino, by $\widetilde{Y}$. Then from Eq. (\ref{mix2}) we have \bea
\label{clZ}\widetilde{X}&=&\cos\psi \widetilde{Z}- \sin\psi \l
\nn \\  \widetilde{Y}&=& \sin\psi \widetilde{Z} + \cos\psi \l
\eea
 The gaugino mass term of the $H_{51}\equiv SU(5)\times U(1)_X$ can be written as:
\bea \label{calL} {\cal L} &\supset& M_5 \l^a_b \l^b_a + M_X \l^2
\eea where $\l^a_b$'s are the gauginos of $SU(5)$
($a,b=1,\ldots,5$ and $\l^a_a=0$)\footnote{referred to by $\widehat{\l^a_b}$ in Eq. \ref{labH51}.}.
After $H_{51}$ is broken to the
SM, with the indices ($\a,\b=1,2,3;r,s=4,5$), we have: \bea
\label{massLH51} {\cal L}_{\text{mass}} &=& M_5
[(\widehat{\l^\a_\b} \widehat{\l^\b_\a})^2 + (\widehat{\l^r_s}
\widehat{\l^s_r})^2 + \widetilde{Z}^2] + M_X \l^2\eea where
$\widehat{\l^\a_\b} =\l^\a_\b - \frac{1}{3} \d^\a_\b \l^\g_\g$ are
the gaugino fields of $SU(3)_C$ (Similarly, $\widehat{\l^r_s}$ are
the $SU(2)_L$ gauginos) and
$\widetilde{Z}^2 = \frac{1}{3} (\l^\a_\a)^2 + \frac{1}{2}
(\l^r_r)^2 $ is the squared  $U(1)_Z$ gaugino field. From
Eq.(\ref{massLH51}) and using Eq.(\ref{clZ}) we get: \bea
\label{mass_rel_H51} \left. M_2\right|_{M_{HB}}=\left. M_3\right|_{M_{HB}}=\left. M_5\right|_{M_{HB}} &,&
\left. M_1\right|_{M_{HB}}=M_5\sin^2\psi+M_X\cos^2\psi= \left.\frac{g_X^2 M_5 + 24 g_5^2
M_X}{g_X^2 + 24 g_5^2} \right|_{M_{HB}}\eea To summarize, we obtained by calculating the mixing of the
two $U(1)$'s the formulae relating the MSSM-guagino masses
($M_1,M_2,M_3$) to the intermediate group $H_{51}$-gaugino
masses ($M_5,M_X$) and the coupling constants, which are valid at
the scale where the breaking of the intermediate group to the SM
occurs.

\section{The RG running and the MSSM gaugino mass ratios}

In section 2, we computed the $H$-gaugino mass ratios at the GUT scale $M_{GUT}$, whereas in section 3
we expressed, at the intermediate breaking scale $M_{HB}$, the MSSM gaugino masses in terms of the $H$-gaugino masses
and the coupling constants. Thus, it is necessary
to introduce the running factors for the gauge couplings of the
intermediate group ($\a_i \equiv \frac{g_i^2}{4 \pi}$) from
$M_{GUT}$ to $M_{HB}$: \bea r_i=\frac{\alpha_i(t)}{\alpha_i(t_0)}
&,& t=\log\frac{M_{GUT}^2}{Q^2} \eea with $Q^2=M_{HB}^2$ and $t_0=0$
corresponding to $Q^2=M_{GUT}^2$, and we assume unification at
$M_{GUT}$ ($\a_i(t_0)=\a$). We define the ratio \bea
\label{running-ratio} R(i,j) \equiv \frac{r_i}{r_j}&=&
\frac{1+\frac{\a}{2\pi} b_j t}{1+\frac{\a}{2\pi} b_i t}\eea with
$b_i$ the beta function coefficients, and use the one-loop
renormalization equations for the evolution of the gaugino masses
and the coupling constants: \bea \label{1loop}
\frac{M_i(t)}{g_i^2(t)} &=& \frac{M_i(t_0)}{g_i^2(t_0)} \eea

With this we can obtain our final results of the MSSM gaugino mass ratios at the intermediate
scale $M_{HB}$ as follows:

\begin{itemize}

\item {\underline{$SO(10)\to G_{422}$ by ${\bf 54}$}}

Eqs.(\ref{mass_rel_g422_first},\ref{mass_rel_g422_second}) and
(\ref{54G422})lead to \bea
\frac{M_2(t)}{M_3(t)}=-\frac{3}{2}R(2_L,4) &,&
\frac{M_1(t)}{M_3(t)}=\frac{-5 R(2_R,4)}{4 R(2_R,4)+6}
\label{m13-54-422} \eea We note that we get the gaugino masses
$M_a$(a=1,2,3) in the ratio $-\frac{1}{2}:-\frac{3}{2}:1$ when the
two scales are equal ($M_{HB}=M_{GUT}$) in accordance with the
results of \cite{martin} obtained via a different approach.
However, it is instructive to notice here that the functional form
of the ratio $M_1/M_3$, in terms of the `RG'-factor $R(2_R,4)$,
 in equation (\ref{m13-54-422}) can not be deduced directly, by simple RG running, from
 its value $(-\frac{1}{2})$ when $R(2_R,4)=1$ corresponding to two equal scales. This
 comes because the mixing of two $U(1)$'s, one from $SU(4)_C$ and the other from $SU(2)_R$, to
 give $U(1)_Y$ happens at the
intermediate scale $M_{HB}$, and use of
Eq.(\ref{mass_rel_g422_second}) is essential in
order to take account of this mixing.

\item{\underline{$SO(10)\to G_{422}$ by ${\bf 210}$}}

Eqs.(\ref{mass_rel_g422_first},\ref{mass_rel_g422_second}) and
(\ref{0},\ref{-1}) lead to \bea M_3(t)=0 &,&
\frac{M_1(t)}{M_2(t)}=\frac{-3}{3+2 R(2_R,4)} \label{m12-210-422}
\eea where the symmetric evolution of $\a_{2R}$ and $\a_{2L}$ puts
$R(2_R,2_L)=1$. This reduces to the `known' value
$\frac{M_1}{M_2}= -\frac{3}{5}$ when $M_{HB}=M_{GUT}$
\cite{martin}. We note that the possibility of gluinos being
massless is not phenomenologically excluded.

\item {\underline{$SO(10)\to G_{422}$ by ${\bf 770}$}}

Eqs.(\ref{mass_rel_g422_first},\ref{mass_rel_g422_second}) and
(\ref{5/2}) lead to \bea \label{m13-770-422}  \frac{M_1(t)}{M_3(t)}=\frac{19
R(2_R,4)}{6+4R(2_R,4)} &,& \frac{M_2(t)}{M_3(t)}=\frac{5}{2}R(2_L,4)
  \eea
We see that when $M_{HB}=M_{GUT}$ the results of the gaugino
masses $M_a$(a=3,2,1) reduce, as expected, to
$1:\frac{5}{2}:\frac{19}{10}$ in ratio \cite{martin}.

\item \underline{$SO(10)\to SU(2)\times SO(7)$ by ${\bf 54}$}

  Eqs. (\ref{m4m7},\ref{mass_rel_so7}) and (\ref{54s03so7}) lead to
  gaugino masses,
at the intermediate scale $M_{HB}$, in the ratio: \bea M_3: M_2:
M_1 &=& 1:-\frac{7}{3} R(2_L,4) :1\eea which reduces to
$1:-\frac{7}{3}:1$  when $M_{HB}=M_{GUT}$ \cite{chamoun}.

\item \underline{$SO(10)\to SU(2)\times SO(7)$ by ${\bf 770}$}

Eqs. (\ref{m4m7},\ref{mass_rel_so7}) and (\ref{7}) lead to \bea
\frac{M_1(t)}{M_3(t)} = 1 &,& \frac{M_2(t)}{M_3(t)}=7 R(2_L,4)
  \eea which reduce respectively to
$1,7$, when $M_{HB}=M_{GUT}$.

\item \underline{ $SO(10)\to H_{51}$ by ${\bf 210}$}

Eqs. (\ref{mass_rel_H51}) and (\ref{-4}) lead to \bea
\frac{M_2(t)}{M_3(t)} = 1 &,& \frac{M_1(t)}{M_3(t)}=\frac{-95
R(1_X,5)}{R(1_X,5)+24} \label{m-210-H51}\eea Again, these
functional forms are consistent with the `known' values of the
gaugino mass $M_a$(a=3,2,1) ratios $1:1:-\frac{19}{5}$ obtained in
\cite{martin} using a different method when $M_{HB}=M_{GUT}$.
However, their values at $M_{GUT}$ and RG running alone are not enough
to deduce the `functional' forms, and one needs to carefully
consider the normalization and mixing of $U(1)_X$ and $U(1)_Z$,
which was done in Eq.(\ref{mass_rel_H51})

\item \underline {$SO(10)\to H_{51}$ by ${\bf 770}$}

Eqs. (\ref{mass_rel_H51}) and (\ref{ratio16}) lead to \bea
\frac{M_2(t)}{M_3(t)}=1 &,&  \frac{M_1(t)}{M_3(t)}= \frac{385
R(1_X,5)}{24+R(1_X,5)}  \eea which reduce respectively to
$1,\frac{77}{5}$ if $M_{HB}=M_{GUT}$, in accordance with
\cite{martin}.

\end{itemize}

We compute now the beta coefficients for the RG running. We shall
consider that the scale $M_{HB}$ is above the threshold of
creating the superpartners of the known particles, so we use the
RG equations of the SUSY-GUT \cite{vaughin}: \bea b_i &=& S_i(R) -
3 C_i(G)\eea with $S_i(R)$ is the Dynkin index of the irrep $R$
summed over all chiral superfields, normalized to $1/2$ for each
fundamental irrep of $SU(N)$, and $C_i(G)$ is the Casimir
invariant (equal to the Dynkin index of the adjoint
representation) which satisfies $C(SU(N)) = N,C(U(1)) = 0$. In
order to single out the Higgs contribution, we write: \bea S_i(R)
&=& F_i + H_i \eea  and we shall assume we have $N_g=3$ families
of fermions which span an $SO(10)$-$\bf 16$ spinor irrep.

As to the Higgs field, we only consider the Higgs field responsible for the breaking of the intermediate group $H$. These
Higgs fields would include the MSSM Higgses but the way in which this is carried out is model-dependent. As to the Higgs fields responsible for the breaking of $SO(10)$, we do not consider them since they get masses of order of $M_{GUT}$, and some
will be `eaten' by the gauge bosons.

As explained in section 3, we need a Higgs field $\Phi$ in an $\bf
16$-irrep of $SO(10)$ in both cases corresponding to $H= G_{422}$
and $H=H_{51}$, whereas we need a Higgs field $\Phi$ in an $\bf
45$-irrep of $SO(10)$ in the case $H=SU(2)\times SO(7)$, whence we
have the table:

\begin{table}[htbp]
\begin{center}
\begin{tabular}{||c||c||c|c|c|c||c|c|c|c||c||}
\hline
   $H$ & Higgs & $F_i$ &  $H_i$ & $C_i$ & $b_i$ & $F_j$ & $H_j$ & $C_j$ & $b_j$ & MSSM\\
\hline
 $G_{422}$& {\bf{16}} & $2$ & $2$ & $2$ & $2$ & $2$ &  $2$
 & $4$ & $-4$ & $b_1^{MSSM}=\frac{33}{5}$\\
\hline
$SU2 \times SO7$& {\bf{45}} & $4$ & $16$ & $2$ & $22$ & $2$ &  $8$
 & $4$ & $2$ & $b_2^{MSSM}=1$\\
\hline
$H_{51}$& {\bf{16}} & $2$ & $2$ & $0$ & $8$ & $2$ &  $2$
 & $5$ & $-7$ & $b_3^{MSSM}=-3$\\
\hline
\hline
\end{tabular}
\end{center}
 \caption{$(i, j)=(2_R, 4)$ or $(2_L, 4)$  for $H= G_{422}$ or $H = SU2 \times SO7$ (broken
 at $M_{GUT}$ to $H^{\prime}=SU2 \times SU4$), whereas $(i,j)=(1_X, 5)$ for $H= H_{51}$.
 We put also the MSSM beta function coefficients.}
\label{beta coefficients}
\end{table}

\begin{landscape}
\begin{table*}[htbp]
\begin{center}
{\small
\begin{tabular}{|c||c||c|c|c|c||c|c|c|c|}
\hline
   Irrep & $H$ & \multicolumn{4}{|c||}{$M_1/M_3 $} &  \multicolumn{4}{|c|}{$M_2/M_3$}\\
\hline
 $M_{HB}=$& &  & $M_{GUT}$ & $10^8$ & $10^{3}$ &  &  $M_{GUT}$
 & $10^8$ & $10^{3}$ \\
\hline
\hline
{\bf 54} & $G_{422}$ & $\frac{-5 R(2_R,4)}{6+4 R(2_R,4)}$ & $-1/2$ & $\begin{matrix} 0.88 \cr (3.21)
 \end{matrix}$ & $\begin{matrix} 2.27 \cr (1.96)
 \end{matrix}$ &
$-\frac{3}{2} R(2_L,4)$ & $-3/2$ & $\begin{matrix} 0.93 \cr (3.13)
 \end{matrix}$ & $\begin{matrix} 1.45 \cr (1.58)
 \end{matrix}$ \\
\hline
 & $SU2 \times SO7$ & $1$ & $1$ & $\begin{matrix} 1 \cr (-6.42)
 \end{matrix}$ & $\begin{matrix} 1 \cr (-3.92)
 \end{matrix}$ &
$- \frac{7}{3} R(2_L,4)$ & $-7/3$ & $\begin{matrix} -0.36 \cr (4.88)
 \end{matrix}$ & $\begin{matrix} -0.31 \cr (2.45)
 \end{matrix}$ \\
\hline
\hline
{\bf 210} & $G_{422}$ & $m=\frac{-3}{3 + 2 R(2_R,4)}$ & $m=-\frac{3}{5}$ &
$\begin{matrix} m=-1.70 \cr (=-1.84)
 \end{matrix}$ & $ \begin{matrix} m=-2.82 \cr (=-2.24)
 \end{matrix}$ &
$\infty$ & $\infty$ & $\infty$ & $\infty$ \\
\hline
 & $H_{51}$ & $\frac{-95 R(1_X,5)}{24 + R(1_X,5)}$ & $-19/5$ & $\begin{matrix} 2.21 \cr (24.38)
 \end{matrix}$ & $\begin{matrix} 2.67 \cr (14.90)
 \end{matrix}$ &
$1$ & $1$ & $\begin{matrix} 1 \cr (-2.09)
 \end{matrix}$ & $\begin{matrix} 1 \cr (-1.05)
 \end{matrix}$ \\
\hline
\hline
{\bf 770} & $G_{422}$ & $\frac{19 R(2_R,4)}{6+4 R(2_R,4)}$ & $19/10$ & $\begin{matrix} -3.34 \cr (-12.19)
 \end{matrix}$ & $\begin{matrix} -8.63 \cr (-7.45)
 \end{matrix}$ &
$\frac{5}{2} R(2_L,4)$ & $5/2$ & $\begin{matrix} -1.55 \cr (-5.22)
 \end{matrix}$ & $\begin{matrix} -2.42 \cr (-2.63)
 \end{matrix}$ \\
\hline
 & $SU2 \times SO7$ & $1$ & $1$ & $\begin{matrix} 1 \cr (-6.42)
 \end{matrix}$ & $\begin{matrix} 1 \cr (-3.92)
 \end{matrix}$ &
$7R(2_L,4)$ & $7$ & $\begin{matrix} 1.09 \cr (-14.63)
 \end{matrix}$ & $\begin{matrix} 0.93 \cr (-7.36)
 \end{matrix}$ \\
\hline
 & $H_{51}$ & $\frac{385 R(1_X,5)}{24 + R(1_X,5)}$ & $77/5$ & $\begin{matrix} -8.95 \cr (-98.80)
 \end{matrix}$ & $\begin{matrix} -10.85 \cr (-60.40)
 \end{matrix}$ &
$1$ & $1$ & $\begin{matrix} 1 \cr (-2.09)
 \end{matrix}$ & $\begin{matrix} 1 \cr (-1.05)
 \end{matrix}$ \\
\hline
%\hline
\end{tabular}
}
\end{center}
 \vspace{-.5cm}\caption{Gaugino mass ratios at intermediate scale $M_{HB}$ in the different cases. To each ratio correspond four columns, the first of which gives the general formula whereas the other three give the result when $M_{HB}$ is taking a specific value. Bracketed values denote the gaugino mass ratios when $M_{HB}=M_{GUT}$ evaluated at the same specific energy scale ($10^3$ or $10^8$ $GeV$) as the case of $M_{HB}\neq M_{GUT}$. The following numerical values are taken: $M_{GUT}=10^{16}$,
 $\alpha = 0.1$. Mass scales are evaluated in $GeV$. The parameter $m$ is equal to $\frac{M_1}{M_2}$.}
\label{ratios}
\end{table*}
\end{landscape}

\section{Summary and Discussion}

We summarize our results in Table \ref{ratios}, where we compute the gaugino mass ratios in the different
cases, using equation (\ref{running-ratio}), with $\a \sim 0.1$, $M_{GUT}=10^{16}$ GeV and we take two values for
the intermediate breaking scale $M_{HB}=10^3, 10^8$ GeV.

In order to illustrate in the table the effect of `successive' breakings, we
have enclosed in brackets the values of the gaugino mass ratios at the specific values  $10^3, 10^8$ GeV, had
 the two breakings occurred at one stage ($M_{HB}=M_{GUT}$), using the MSSM running from $E=M_{GUT}$ to $E=10^3\text{ or }10^8$ GeV: \bea  \frac{M_i}{M_j}(E) &=& \frac{M_i}{M_j}(M_{GUT}) \frac{1+\frac{\a}{2\pi}tb^{MSSM}_i}{1+\frac{\a}{2\pi}tb^{MSSM}_j} \eea
where $t=\log(\frac{M_{GUT}}{E})^2$.

We see that gaugino mass ratios, evaluated at the same energy scale, change
 significantly when the intermediate scale is low (say, $10^8$ GeV
or Tev) compared to when the two breaking scales are approximately equal.

We note here that we did not consider the impact of the intermediate scale on
gauge coupling unification for the values of the parameters used
in the table. To check that this unification requirement can be achieved in a way consistent with the low scale experimental measurements would involve model
building details, where one constructs a complete SUSY GUT
model with a full superpotential explicitly written, and in which the gauge coupling unification is realized in two steps of breaking: a task beyond the scope of the work in this paper which does not entail model building particularities.

Having said this though, one should notice that from a phenomenological point of view there
is a more reasonable way to obtain the gaugino
mass ratios at the intermediate scale $M_{HB}$. In fact, once
we fix the partially unified intermediate gauge group $H$ and the intermediate mass scale $M_{HB}$,
the values of
the gauge couplings at $M_{HB}$ can be calculated from the weak scale
data by using RG equations, and then one can use the formulae of the past section to compute the corresponding gaugino mass ratios
assuming gauge coupling unification at $M_{GUT}$.
However, whether or not the numerical values
of the running gauge couplings at a `low' intermediate scale
$M_{HB}$ \footnote{By `low' we mean a scale smaller than $\sim 10^{12} GeV$, so that to be capable of explaining the
smallness of neutrino masses.}, which are necessary to evaluate the gaugino
mass ratios at this scale, can match with the SM gauge couplings measured at the electroweak scale
$M_Z$, provided we insist on having just MSSM between
$M_{HB}$ and  $M_Z$ \footnote{More precisely,
one has MSSM between $M_{HB}$ and $M_S$, the SUSY scale, %(which
%will be chosen as 1 TeV in the numerical calculations)
and SM between $M_S$ and $M_Z$.}, would depend heavily on the
nature of $H$. For instance, if $H=SU(5)\times U(1)$, it is
difficult to get a low intermediate mass scale and unify both
coupling constants to one corresponding to $SO(10)$~\cite{hll}.
Nonetheless, if $H=G_{3221}\equiv SU(3)_C\times SU(2)_L\times
SU(2)_R\times U(1)_{B-L}$, the low intermediate mass scale can be
obtained~\cite{majee}.

As an illustrative example, let us take the case of $H=G_{422}$
and calculate the gaugino mass ratios by way of computing the
values of the gauge couplings at $M_{HB}$ from the weak scale
data, and assuming gauge coupling unification at $M_{GUT}$ (which
can be realized by, say, adding some particle content near
$M_{HB}$ similar to that in \cite{majee}\footnote{In \cite{majee},
with the intermediate group $G_{3221}$ and additional light
supermultiplets with masses around the intermediate mass scale
$M_R$ (corresponding to $M_{HB}$ in the present paper), one could,
within SUSY SO(10) GUT, achieve low values for $M_R$
($10^4-10^{10} GeV$) with $M_{GUT}\sim 10^{16} GeV$.}). With the
numerical values \cite{pdg} ($M_Z = 91.18$ GeV, $\a_S(M_Z) \sim
0.1187$, $\sin^2\th_W \sim 0.2312,\a_{em}^{-1}(M_Z)=127.9 \Rightarrow
\a_{2L}^{-1}(M_Z) = 29.57$ and $\a^{-1}_Y(M_Z) = 58.99$) and the
MSSM beta coefficients from Table \ref{beta coefficients}, we get,
for $M_{HB}=10^4 GeV$, the values: $\a_S^{-1}= 12.91,
\a_{2L}^{-1}= 28.07, \a_Y^{-1} = 49.12$ at $M_{HB}$. Because $H$
breaks into the SM at $M_{HB}$, we have $g_4(M_{HB}) =
g_S(M_{HB})$ and $g_{2R}(M_{HB})= g_{2L}(M_{HB})$. Applying Eqs.
\ref{m13-54-422}, \ref{m12-210-422} and \ref{m13-770-422}, we get
the numerical results of the gaugino mass ratios and show them in
Table \ref{specific model}.

\begin{table}[htbp]
\begin{center}
\begin{tabular}{||c||c|c||}
\hline
   Irrep &  $M_1/M_3$ &  $M_2/M_3$ \\
\hline
{\bf{54}} & $-0.29$ & $-0.68$ \\
\hline
{\bf{210}} & $\infty$ & $M_1/M_2 = -0.76$ \\
\hline
{\bf{770}} & $1.11$ & $1.15$ \\
\hline
\hline
\end{tabular}
\end{center}
 \caption{Gaugino mass ratios at $M_{HB} \sim 10$ TeV for an intermediate
$G_{422}$ group, obtained by computing the values of
gauge couplings at $M_{HB}$ starting from the weak scale data.}
\label{specific model}
\end{table}

In general, considering other models and other intermediate groups, one can say that although some model complexifications might affect the coupling constants evolution,
and consequently the values of the derived gaugino mass
ratios, however the conclusion concerning the significant influence of
the existence of multi-stages in the breaking chain would remain unchanged. The derived mass ratios would be reflected in the electroweak energy scale measurements due to take place in the near future experiments, like the LHC, with interesting phenomenological consequences.

\vspace{0.1cm} {\it{Acknowledgements:}} This work was supported in
part by the National Natural Science Foundation of China under
nos.90503002, 10821504, 10805018 and 10975171. N.C. thanks
CBPF-Brazil, where part of the work has been done, for its
hospitality and
 acknowledges support from TWAS.

\end{document}